\documentclass[twocolumn,english,aps,preprintnumbers,amsmath,amssymb,superscriptaddress]{revtex4}
\usepackage[latin9]{inputenc}
\usepackage{amsmath}
\usepackage{graphicx}

\makeatletter
\@ifundefined{textcolor}{}
{%
 \definecolor{BLACK}{gray}{0}
 \definecolor{WHITE}{gray}{1}
 \definecolor{RED}{rgb}{1,0,0}
 \definecolor{GREEN}{rgb}{0,1,0}
 \definecolor{BLUE}{rgb}{0,0,1}
 \definecolor{CYAN}{cmyk}{1,0,0,0}
 \definecolor{MAGENTA}{cmyk}{0,1,0,0}
 \definecolor{YELLOW}{cmyk}{0,0,1,0}
}

\usepackage{dcolumn}
\usepackage{bm}
\usepackage{color}
\usepackage[final]{pdfpages}

\makeatother

\usepackage{babel}

\begin{document}

\preprint{preprint(\today)}

\title{Putative helimagnetic phase in the kagome metal Co$_{3}$Sn$_{2-x}$In$_{x}$S$_{2}$}

\author{Z.~Guguchia}
\email{zurab.guguchia@psi.ch} 
\affiliation{Laboratory for Muon Spin Spectroscopy, Paul Scherrer Institute, CH-5232 Villigen PSI, Switzerland}

\author{H. Zhou}
\affiliation{International Center for Quantum Materials and School of Physics, Peking University, Beijing, China.}
\affiliation{CAS Center for Excellence in Topological Quantum Computation, University of Chinese Academy of Science, Beijing, China.}

\author{C.N.~Wang}
\affiliation{Laboratory for Muon Spin Spectroscopy, Paul Scherrer Institute, CH-5232 Villigen PSI, Switzerland}

\author{J.-X.~Yin}
\affiliation{Laboratory for Topological Quantum Matter and Advanced Spectroscopy, Department of Physics, Princeton University, Princeton, New Jersey 08544, USA}

\author{C.~Mielke III}
\affiliation{Laboratory for Muon Spin Spectroscopy, Paul Scherrer Institute, CH-5232 Villigen PSI, Switzerland}
\affiliation{Department of Physics, University of Z\"{u}rich, Winterthurerstrasse 190, Z\"{u}rich, Switzerland}

\author{S.S.~Tsirkin}
\affiliation{Department of Physics, University of Z\"{u}rich, Winterthurerstrasse 190, Z\"{u}rich, Switzerland}

\author{I. Belopolski}
\affiliation{Laboratory for Topological Quantum Matter and Advanced Spectroscopy, Department of Physics, Princeton University, Princeton, New Jersey 08544, USA}

\author{S.-S. Zhang}
\affiliation{Laboratory for Topological Quantum Matter and Advanced Spectroscopy, Department of Physics, Princeton University, Princeton, New Jersey 08544, USA}

\author{T.A. Cochran}
\affiliation{Laboratory for Topological Quantum Matter and Advanced Spectroscopy, Department of Physics, Princeton University, Princeton, New Jersey 08544, USA}

\author{T.~Neupert}
\affiliation{Department of Physics, University of Z\"{u}rich, Winterthurerstrasse 190, Z\"{u}rich, Switzerland}

\author{R.~Khasanov}
\affiliation{Laboratory for Muon Spin Spectroscopy, Paul Scherrer Institute, CH-5232
Villigen PSI, Switzerland}

\author{A.~Amato}
\affiliation{Laboratory for Muon Spin Spectroscopy, Paul Scherrer Institute, CH-5232
Villigen PSI, Switzerland}

\author{S.~Jia}
\affiliation{International Center for Quantum Materials and School of Physics, Peking University, Beijing, China.}
\affiliation{CAS Center for Excellence in Topological Quantum Computation, University of Chinese Academy of Science, Beijing, China.}

\author{M.Z.~Hasan}
\affiliation{Laboratory for Topological Quantum Matter and Advanced Spectroscopy, Department of Physics, Princeton University, Princeton, New Jersey 08544, USA}
\affiliation{Lawrence Berkeley National Laboratory, Berkeley, CA, USA}

\author{H.~Luetkens}
\affiliation{Laboratory for Muon Spin Spectroscopy, Paul Scherrer Institute, CH-5232 Villigen PSI, Switzerland}

\begin{abstract}

\textbf{The exploration of topological electronic phases that result from strong electronic correlations is a frontier in condensed matter physics \cite{Keimer,WangZhang,HasanKane,Wen}. One class of systems that is currently emerging as a platform for such studies are so-called kagome magnets based on transition metals. Using muon spin-rotation, we explore magnetic correlations in the kagome magnet Co$_{3}$Sn$_{2-x}$In$_{x}$S$_{2}$ as a function of In-doping, providing putative evidence for an intriguing incommensurate helimagnetic (HM) state. Our results show that, while the undoped sample exhibits an out-of-plane ferromagnetic (FM) ground state, at 5 ${\%}$ of In-doping the system enters a state in which FM and  in-plane antiferromagnetic (AFM) phases coexist. At higher doping, a HM state emerges and becomes dominant at the critical doping level of only $x_{\rm cr,1}$ ${\simeq}$ 0.3. This indicates a zero temperature first order quantum phase transition from the FM, through a mixed state, to a helical phase at $x_{\rm cr,1}$. In addition, at $x_{\rm cr,2}$ ${\simeq}$ 1, a zero temperature second order phase transition from helical to paramagnetic phase is observed, evidencing a HM quantum critical point (QCP) in the phase diagram of the topological magnet Co$_{3}$Sn$_{2-x}$In$_{x}$S$_{2}$. The observed diversity of interactions in the magnetic kagome lattice drives non-monotonous variations of the topological Hall response of this system.}

\end{abstract}
\maketitle

\section{Introduction}

Kagome lattice systems are an ideal setting in which strongly correlated topological electronic states may emerge \cite{Keimer,WangZhang,HasanKane,Wen,JXYin2,GuguchiaNat,Mazin,LYe,THan,JXYin1,Legendre,Ghimire}. The material Co$_{3}$Sn$_{2}$S$_{2}$ has a layered crystal structure featuring a CoSn kagome lattice and was shown to have an out-of plane ferromagnetic ground state (Curie temperature of $T_{\rm C}$ ${\simeq}$ 177 K) with a magnetization arising mainly from the cobalt moments. This ferromagnetic semimetal has been shown to possess both a large anomalous Hall conductivity of 1130 ${\Omega}$$^{-1}$cm$^{-1}$ \cite{FelserCSS,Huibin} and a giant anomalous Hall angle of 20${\%}$ \cite{FelserCSS,Muechler,Wang} in the 3D bulk. Density functional theory (DFT) calculations have predicted three pairs of Weyl points in one Brillouin zone located only 60 meV above the Fermi level \cite{QXu}. Recently, by combining ARPES and ${\mu}$SR experiments we established Co$_{3}$Sn$_{2}$S$_{2}$ as a material that hosts a topologically non-trivial band structure and frustrated magnetism \cite{GuguchiaNat,Belopolski}. Namely, we showed that the Co spins have both ferromagnetic interactions along the $c$-axis (Fig. 4d) and antiferromagnetic interactions (Fig. 4e) within the kagome plane, and that there is a temperature dependent and volume wise competition between these two ordering tendencies. Moreover, we find a near-perfect correlation between the anomalous Hall conductivity and the ferromagnetic volume fraction as a function of temperature. Hund`s coupling along the $z$-direction, Mott physics and electron-mediated interactions between the half-filled orbitals was shown \cite{Legendre} to reproduce the out-of-plane ferromagnetism and an antiferromagnetic transition with a 120$^{\circ}$ spin ordering in the $xy$ plane, as we observed experimentally \cite{GuguchiaNat}. Frustration in the magnetic kagome lattice was also shown to drive an exchange biased anomalous Hall effect \cite{Lachman}. However, despite knowing the remarkable thermodynamic response of the magnetic and topological states in Co$_{3}$Sn$_{2}$S$_{2}$, a possible quantum phase transition (QPT), as shown from standard magnetisation measurements \cite{KassemPhd,Huibin}, has received less attention, despite being vital for a full understanding of the link between magnetic and topological transitions in this system. Unlike ordinary phase transitions driven by thermal energy, a QPT is characterized at a particular value of a non-thermal parameter (e.g., doping or pressure) where a phase transition takes place between a quantum ordered phase and a quantum disordered phase at zero temperature \cite{Chubukov,Sachdev,Frandsen2016,GuguchiaPRM,UemuraMnSi,Goko}. Since the bulk magnetization \cite{Huibin} is a very indirect probe for the magnetic structure, it is essential to explore the temperature-doping phase diagram using a microscopic magnetic probe. Namely, to study how the phase competition between the AFM state and the topological FM structure (which exhibits a strong AHC response in undoped Co$_{3}$Sn$_{2}$S$_{2}$) evolves as a function of In-doping. Furthermore, understanding the nature of the magnetic to paramagnetic quantum phase transition (QPT) is crucial, e.g., whether it is first- or second-order (continuous) in nature. A second-order QPT leads to quantum criticality that may engender unusual properties and novel electronic phases. For a first-order QPT  \cite{Chubukov,Frandsen2016,GuguchiaPRM,UemuraMnSi,Goko}, the system would be expected to exhibit behaviour such as phase coexistence and abrupt changes in the ground state, not necessarily manifesting quantum criticality in the same way. 

\begin{figure*}[t!]
\includegraphics[width=0.9\linewidth]{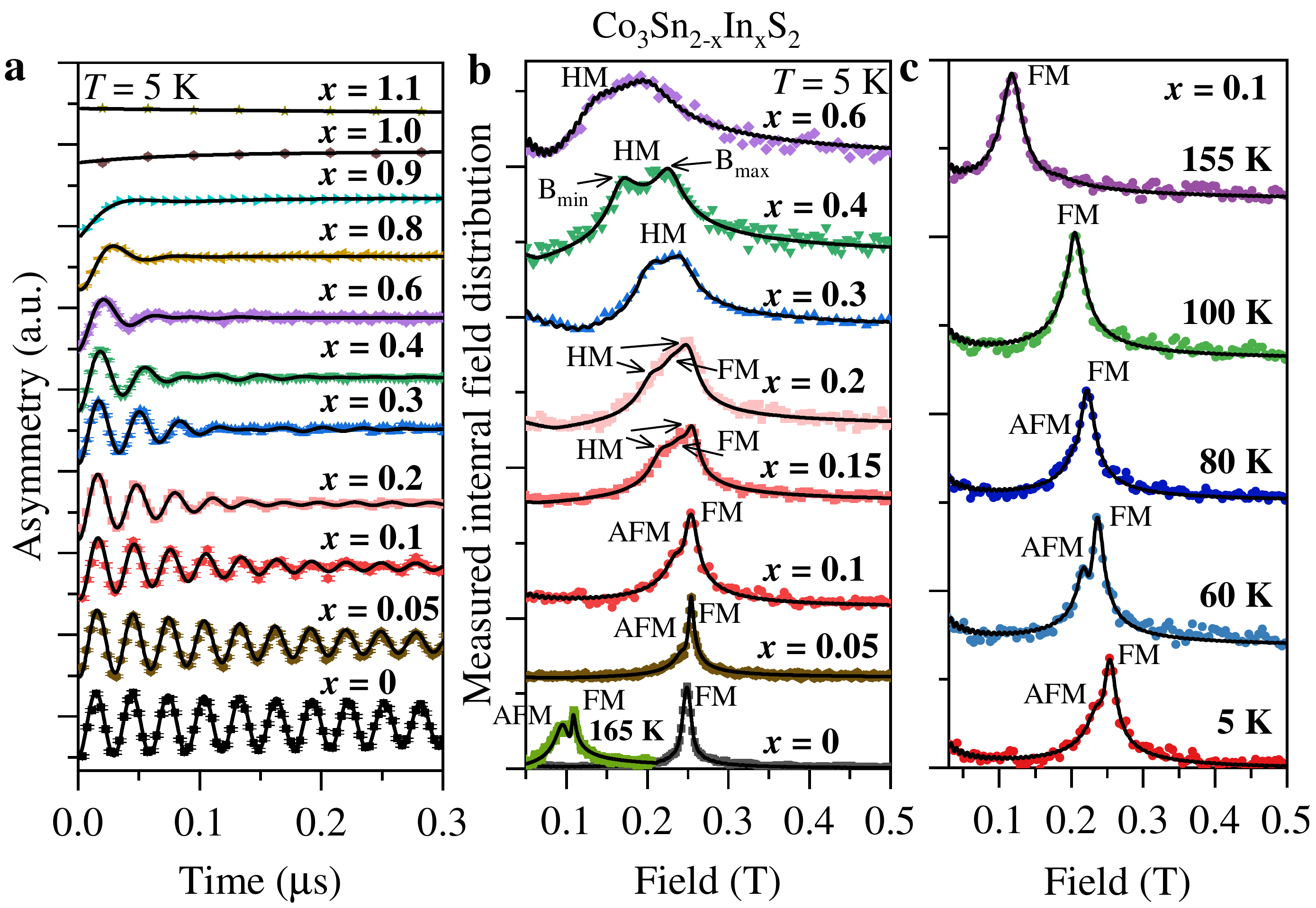}
\vspace{-0.1cm}
\caption{ (Color online) \textbf{Zero-field (ZF) ${\mu}$SR time spectra for CoSn$_{2-x}$In$_{x}$S$_{2}$.}
Zero-field (ZF) ${\mu}$SR time spectra (a) and the corresponding Fourier transform amplitudes (b) of the oscillating components for undoped and various In-doped CoSn$_{2-x}$In$_{x}$S$_{2}$, recorded at 5 K. FFT amplitude for the $x$ = 0 sample, recorded at 165 K, is also shown. (c) Fourier transform amplitudes of the oscillating components of the ${\mu}$SR time spectra for various temperatures for the $x$ = 0.1 sample. FFT amplitudes were normalised to its maximum value to make the signals well visible especially for samples for high In-doping. FM, AFM and Hel denote the out-of-plane ferromagnetic, in-plane antiferromagnetic and helical orders, respectively (see the text for the explanation).}  
\label{fig1}
\end{figure*}

 We have carried out high-resolution muon spin relaxation/rotation (${\mu}$SR) and transport experiments to systematically characterize the temperature-doping phase diagram in Co$_{3}$Sn$_{2-x}$In$_{x}$S$_{2}$. As a highly sensitive probe of local magnetism, ${\mu}$SR can measure the local order parameter and magnetically ordered volume fraction independently, and is therefore ideally suited to determine the first- or second-order behaviour at a QPT. This makes ${\mu}$SR highly complementary with probes of long-range magnetic order such as neutron diffraction. The results presented here demonstrate that the out-of-plane FM ground state transitions into mixed (FM+AFM)- or  (FM+HM)-states 5 ${\%}$ and 10 ${\%}$ of In-doping, respectively. Finally, the putative HM state is established at the critical doping level of $x_{\rm cr,1}$ ${\simeq}$ 0.3, above which the intrinsic anomalous Hall conductivity becomes negligibly small. Interestingly, the QPT between the FM and helical states is first order. In contrast, the quantum evolution of the helimagnet-paramagnet, which is accompanied by a metal-semiconductor transition at zero temperature happens as a second-order transition with $x_{\rm cr,2}$ ${\simeq}$ 1. These results demonstrate not only the effective In-doping tuning of the competition between FM and AFM magnetic correlations, but that the competing magnetic correlations determine the anomalous Hall response of this system. Moreover, the current results uncover the novel first-order QPT transition as well as an AFM quantum critical point (QCP) in the phase diagram of the kagome magnet Co$_{3}$Sn$_{2-x}$In$_{x}$S$_{2}$.  

\begin{figure*}[t!]
\includegraphics[width=1.0\linewidth]{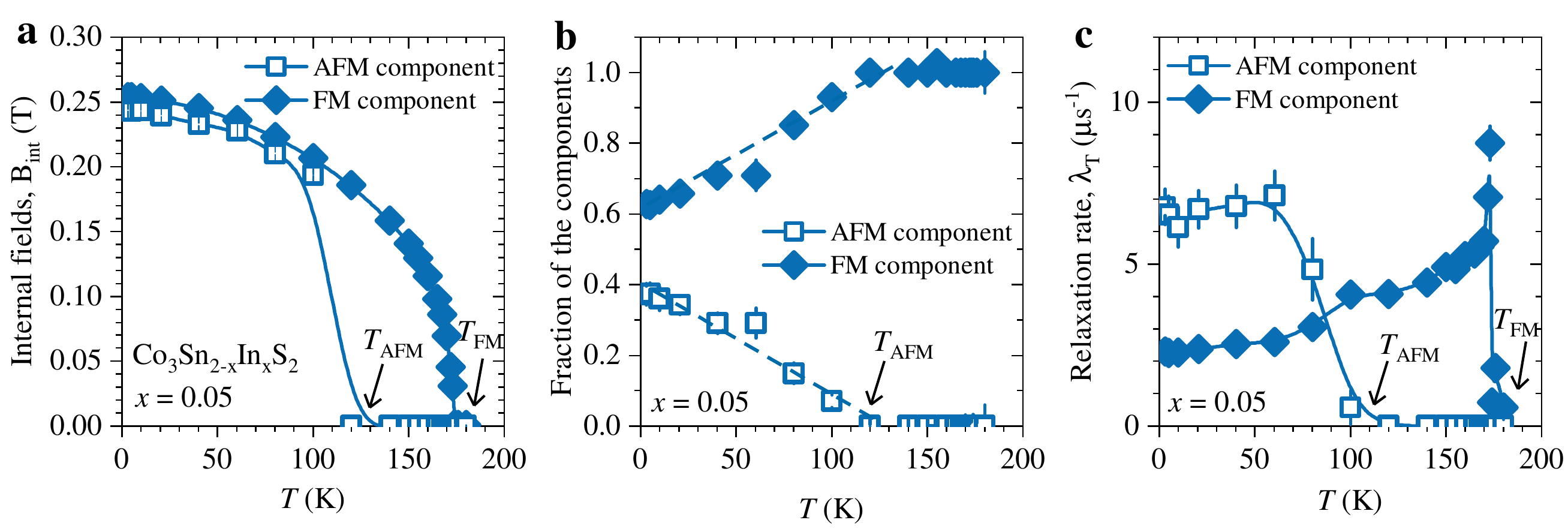}
\includegraphics[width=1.0\linewidth]{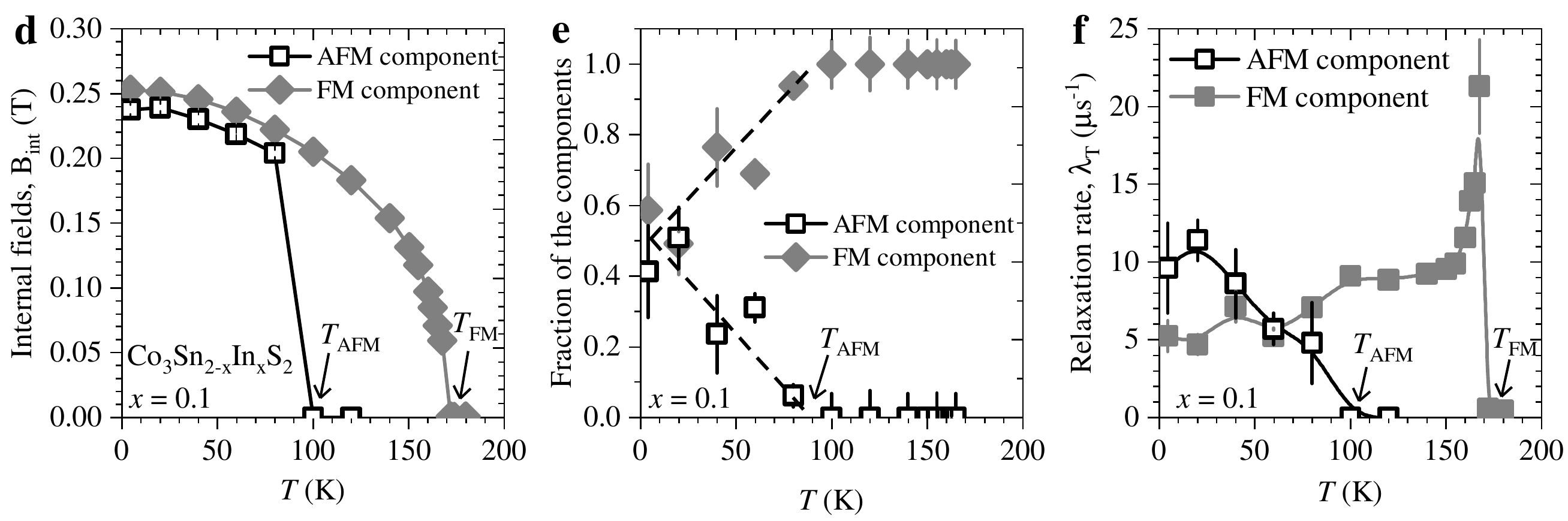}
\vspace{-0.7cm}
\caption{ (Color online) \textbf{The temperature dependence of various fit parameters of the ${\mu}$SR experiments.}
The temperature dependences of the internal magnetic fields (a,d), the relative volume fractions (b,e) and the transverse relaxation rates (c,f) of the two magnetically ordered regions in the samples $x$ = 0.05 and 0.1. Arrows mark the critical temperatures $T_{\rm AFM}$ and $T_{\rm FM}$ for AFM and FM components.} 
\label{fig1}
\end{figure*}

\section{Results and discussion}

Fig. 1(a) displays representative zero-field (ZF) ${\mu}$SR time spectra for Co$_{3}$Sn$_{2-x}$In$_{x}$S$_{2}$ taken at 5 K for different indium dopant concentrations. The ZF oscillation frequency and amplitude are directly proportional to the size of the internal magnetic field and therefore the size of the ordered moment and the magnetically ordered volume fraction, respectively. ZF ${\mu}$SR spectra for Co$_{3}$Sn$_{2-x}$In$_{x}$S$_{2}$ reveal clear oscillations for In-doping levels $x$ ${\leq}$ 0.9. This indicates the existence of a well-defined internal field, as expected in the case of long-range magnetic order. For the $x$ = 1 sample, we observe relaxation occurring only in a small fraction of the signal with no well-defined oscillations. Longitudinal field ${\mu}$SR  (see Supplementary Figures S6 and S7) experiments indicate that this relaxation arises from dynamic spin fluctuations from a small volume fraction of the sample. The majority of the fraction is in the paramagnetic state as evidenced by the weak ${\mu}$SR depolarization and its Gaussian functional form arising from the interaction between the muon spin and randomly oriented nuclear magnetic moments \cite{Toyabe} (see Supplementary Figure S6 and the corresponding text). For $x$ = 1.1, the entire sample is in the paramagnetic state. To better visualize the effect of doping on the magnetic response, we show the Fourier transform amplitudes of the oscillating components of the ${\mu}$SR time spectra at $T$ = 5 K as a function of In-doping (Fig. 1b), which is a measure for the probability distribution of internal fields sensed by the muon ensemble. For the undoped sample, the FFT amplitude recorded at 165 K is also shown. For $x$ = 0 sample, a well-defined single internal field is observed with a narrow width of the field distribution in the sample, which was  explained by the presence of a homogeneous out-of-plane ferromagnetic ground state \cite{GuguchiaNat} (Fig. 4d, although some canting away from the $c$-axis is possible). However, at temperatures above $T_{\rm C}^{*}$ ${\sim}$ 90 K, a second lower internal field is observed in the ${\mu}$SR spectra (165 K spectrum is shown as an example), which was interpreted as the appearance of the high-temperature in-plane AFM state (see Fig. 4e). It was also shown that an AFM component with a larger width of the field distribution (component II) develops at the cost of the FM (component I) component, since the appearance of the in-plane AFM component above  $T_{\rm C}^{*}$ ${\sim}$ 90 K is accompanied by the reduction of the volume fraction of the FM one \cite{GuguchiaNat}. It is interesting that for the low In-doped $x$ = 0.05 and 0.1 samples, two distinct internal fields appear in the ${\mu}$SR spectra already at the base temperature. This suggests the presence of spatially separated antiferromagnetically and ferromagnetically ordered regions already in the ground state of the $x$ = 0.05 and 0.1 samples, implying that $T_{\rm C}^{*}$ ${\simeq}$ 0. But, as the temperature increases, a transition from a two-field to a single-field is observed across 100 K (see Figure 1c) and only the high-field narrow FM component remains at high temperatures. So, the results of $x$ = 0.05 and 0.1 samples show that the FM component becomes dominant and occupies the full sample volume at high temperatures, while the ground state is characterised by spatially separated FM and AFM regions. For the samples with $x$ ${\geq}$ 0.3, we observed the field distribution (Eq. 3) which is characterised by minimum ($B_{\rm min}$) and maximum ($B_{\rm max}$) cutoff fields (see Fig. 2b), which is consistent with the incommensurate helimagnetic order, observed in MnSi \cite{AmatoPRB,Schenck,Yaouanc} and MnP
\cite{KhasanovPRBrapid,KhasanovJPCM}. The cutoff fields $B_{\rm min}$ and $B_{\rm max}$ are measures of the magnetic order parameter of the helimagnetic state. The difference between $B_{\rm min}$ and $B_{\rm max}$ increases with increasing In-content. For $x$ = 0.15 and 0.2 samples, both the commensurate FM peak and incommensurate HM field distribution are observed simultaneously. These results show that within a narrow region of In-concentration, the magnetic ground state changes from FM to a mixed FM and AFM/HM and finally to a fully HM state.

\begin{figure*}[t!]
\includegraphics[width=1.0\linewidth]{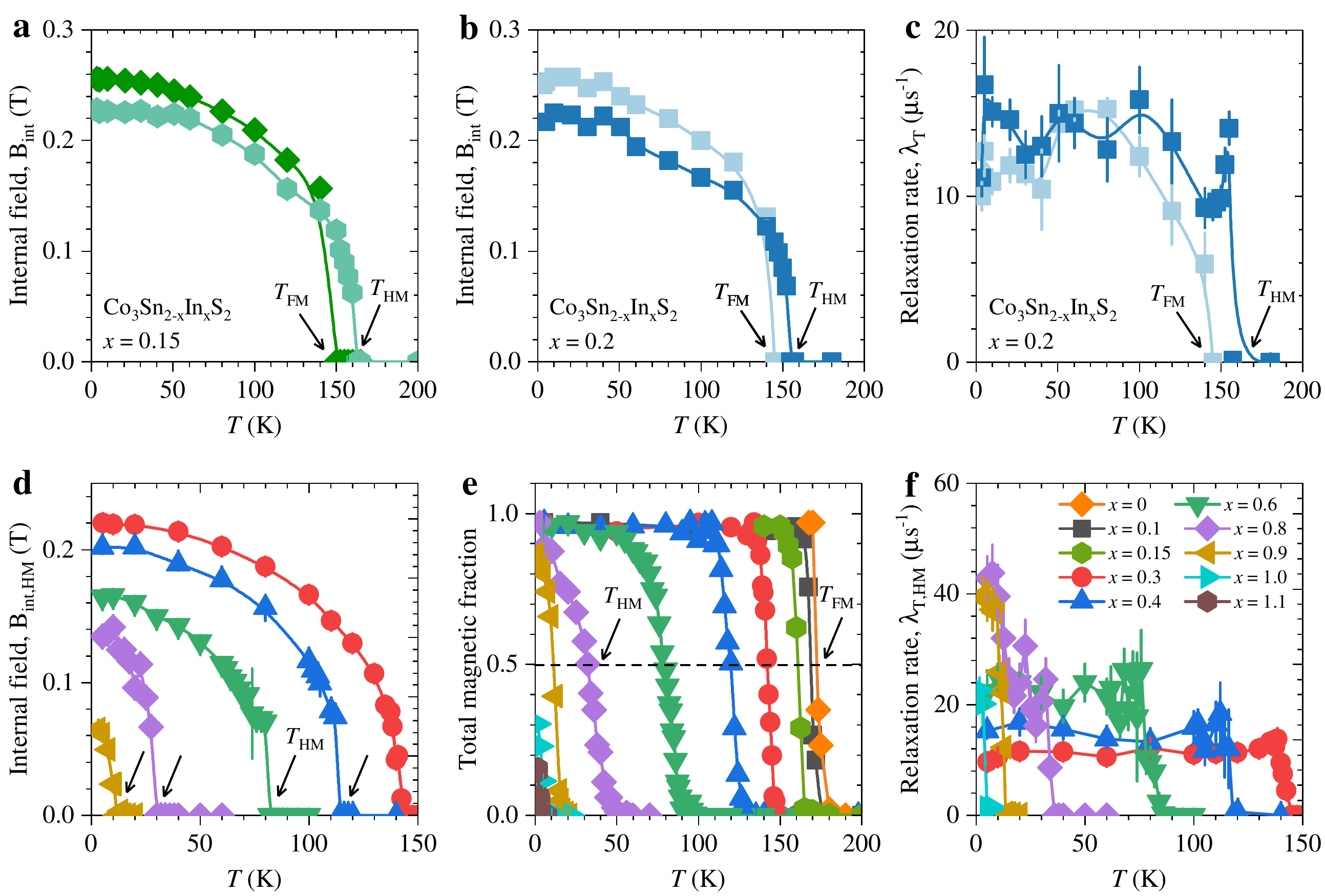}
\vspace{-0.7cm}
\caption{ (Color online) \textbf{The temperature dependence of various fit parameters of the ${\mu}$SR experiments.}
The temperature dependences of the internal magnetic fields of the two magnetically ordered regions in the samples $x$ = 0.15 (a) and 0.2 (b). (c) The temperature dependence of the transverse relaxation rates of the two magnetically ordered regions in the sample $x$ = 0.2. Arrows mark the critical temperatures $T_{\rm FM}$ and $T_{\rm Hel}$ for FM and helical magnetic components, respectively.
The temperature dependences of the average value of magnetic fields (d) and the transverse relaxation rates (f) for the helical magnetic component shown for various In-doped samples. (e) The temperature dependence of the total magnetic volume fraction for various In-doped samples. Arrows in panel (e) mark the temperature at which the magnetic fraction is 50${\%}$.} 
\label{fig1}
\end{figure*}

Using the data analysis procedure described in the methods sections, we obtained quantitative information about the muon spin relaxation rate, the magnetic volume fraction, and the magnitude of the static internal field at the muon site. The temperature dependences of the internal fields ($B_{\rm int}$ = ${\omega}$/${\gamma}_{\mu}$$^{-1}$) for the two components for $x$ = 0.05 and 0.1 samples are shown in the Figs. 2a and d, respectively. Both order parameters show a monotonous decrease. It is important to note that the two components have significantly different transition temperatures, with the AFM component having an onset at $T_{\rm {AFM}}$ ${\simeq}$ 120 K ($x$ = 0.05) and 100 K ($x$ = 0.1), and the FM component at $T_{\rm {FM}}$ ${\simeq}$ 175 K ($x$ = 0.05) and 170 K ($x$ = 0.1). Remarkably, there is a volume-wise interplay between FM and AFM states similar to that observed in the undoped sample, since the volume fraction of the AFM component decreases with increasing temperature and above 120 K and 100 K the full volume of the samples $x$ = 0.05 and 0.1 becomes ferromagnetically ordered (see Figure 2b and e). So, the value of the AFM order temperatures $T_{\rm {AFM}}$ for $x$ = 0.05 and 0.1 samples are reduced significantly compared to the $x$ = 0 sample. However, the AFM state extends down to the base-$T$ unlike in the undoped sample \cite{GuguchiaNat}. Figures 2c and f show the temperature dependences of the transverse depolarization rates ${\lambda}_{\rm T,AFM}$ and ${\lambda}_{\rm T,FM}$ for the $x$ = 0.05 and 0.1 samples for the AFM and the FM components, respectively. In general, muon spin depolarization may be caused by the finite width of the static internal magnetic field distribution and/or by dynamic spin fluctuations. In the present case, the damping is predominantly due to the former effect, as the latter possibility is excluded by the small value of the longitudinal relaxation rate ${\lambda}_{\rm L}$. 
The value of ${\lambda}_{\rm T,AFM}$ is at least a factor of two higher than ${\lambda}_{\rm T,FM}$, which is similar to the undoped sample, indicating a more disordered nature of the AFM state in Co$_{3}$Sn$_{2-x}$In$_{x}$S$_{2}$. ${\lambda}_{\rm T,AFM}$ monotonously decreases with increasing temperature, similar to the internal field and becomes zero at $T_{\rm {AFM}}$. On the other hand, ${\lambda}_{\rm T,FM}$ increases with increasing temperature and shows a maximum near the transition temperature $T_{\rm {FM}}$ with a higher value as compared to the one at 5 K. Such an unconventional temperature dependence of the depolarization rate was also observed in the $x$ = 0 sample \cite{GuguchiaNat} and implies that the magnetism becomes more disordered upon approaching the transition. The domain wall motion near the critical temperature, as recently observed with imagery analysis \cite{Sandeep}, could justify such a behaviour. Figures 3a and b show the temperature dependences of the internal fields for the FM and HM components for $x$ = 0.15 and 0.2 samples, respectively. We note that for simplicity we define the internal field in the helical phase $B_{\rm int,Hel}$ as the average value of the minimum and the maximum cutoff fields $B_{\rm int,Hel}$ = ($B_{\rm min}$ + $B_{\rm max}$)/2. Both order parameters show a monotonous decrease. Both FM and HM regions exhibit similar transition temperatures. However, the FM fraction is only 30 ${\%}$ and 15 ${\%}$ for $x$ = 0.15 and 0.2 samples, respectively. For $x$ ${\geq}$ 0.3, only HM order is observed and the temperature dependences of the corresponding internal fields $B_{\rm int,Hel}$ are displayed in Figure 3d, revealing a gradual increase in the internal field as the temperature is lowered. This suggests a second-order-like behaviour of the thermal phase transition and reveals the continuous development of the ordered moment size, as is the case for the undoped sample \cite{GuguchiaNat}. The value of $B_{\rm int,Hel}$ at the lowest measured temperature of $T$ = 2 K is continuously reduced with In-doping. $B_{\rm int,Hel}$ becomes zero at the critical doping of $x$ = 1, which is accompanied by a metal to semiconductor transition \cite{Huibin}. The temperature dependence of the magnetic volume fraction extracted from the weak-TF data for all measured In-doped samples is displayed in Fig. 3e. Interestingly, the samples with $x$ ${\leq}$ 0.9, in which the long-range magnetism is observed, are fully ordered at low temperatures. The semiconducting sample $x$ = 1 shows a significantly reduced ordered volume fraction, which is suppressed nearly to zero for $x$ = 1.1. To characterize the changes in magnetic critical temperature, we identify the temperatures $T_{\rm FM}$, $T_{\rm AFM}$, and $T_{\rm HM}$ at which the internal fields of the FM, AFM and HM components start to appear, which coincides with the temperature at which the total magnetic fraction is 50 ${\%}$. 

\begin{figure*}[t!]
\includegraphics[width=1.0\linewidth]{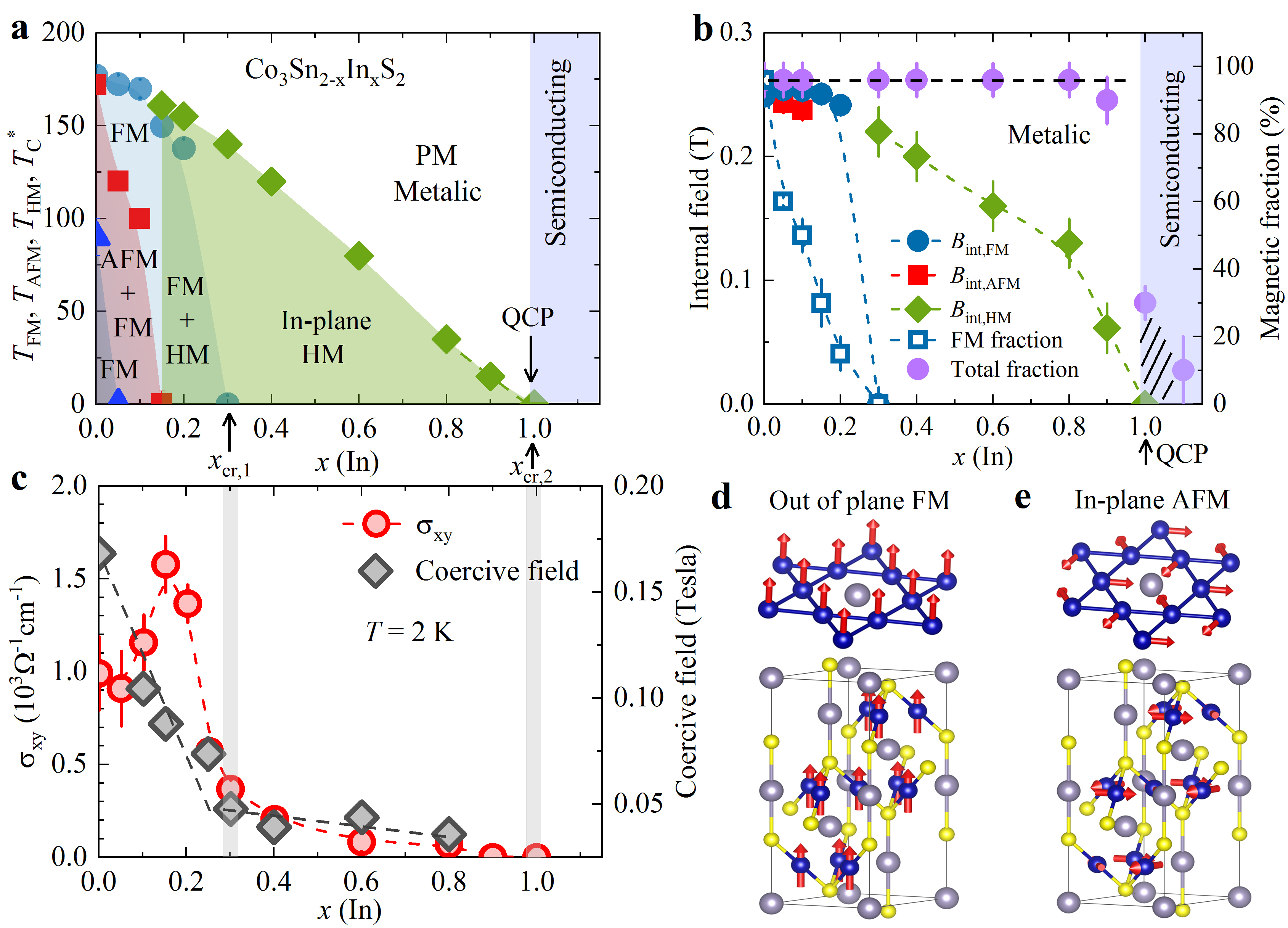} 
\vspace{-0.5cm}
\caption{ (Color online) \textbf{Phase diagrams for CoSn$_{2-x}$In$_{x}$S$_{2}$.}
(a) Temperature-doping phase diagram. $T_{\rm FM}$, $T_{\rm AFM}$ and $T_{\rm HM}$ indicates the FM, AFM and putative HM transition temperatures. $T_{\rm C}^{*}$ denotes the temperature below which only the FM state exists. (b) The doping dependence of the base-temperature value of the internal internal magnetic fields, the FM ordered fraction and the total magnetic fraction. The black dashed line marks the region with weak disordered magnetic response. (c) The In-doping dependence of the in-plane anomalous Hall conductivity ${\sigma}_{xy}$ and the coercive field at $T$ = 2 K. (d-e) Spin structures of Co$_{3}$Sn$_{2}$S$_{2}$, i.e. the $c$-axis aligned FM structure (R-3m$'$ subgroup)  and the in-plane AFM structure (R-3m subgroup). HM state is most likely a helical variation of the in-plane AFM structure.}   
\label{fig1}
\end{figure*}

The results are summarised in Figure 4, showing a complex temperature-doping phase diagram for Co$_{3}$Sn$_{2-x}$In$_{x}$S$_{2}$. Figure 4a displays the evolution of the FM, AFM and HM ordering temperatures as a function of In-content. 
In our previous work \cite{GuguchiaNat}, based on the theoretical group analysis and the local field simulations at the muon site, we showed that the $x$ = 0 sample exhibits a phase separation between in-plane AFM (Fig. 4d) and out-of-plane FM states (Fig. 4e) in the temperature range between $T_{\rm C}^{*}$ ${\simeq}$ 90 K and $T_{\rm AFM}$ ${\simeq}$ 172 K, while below $T_{\rm C}^{*}$ ${\simeq}$ 90 K a homogeneous FM structure is observed. Here, we uncover an incommensurate HM phase and in the temperature-doping phase diagram, we identify five different magnetic phases: a ferromagnetic (FM), phase separation between ferromagnetic and antiferromagnetic (FM+AFM), phase separation between 
ferromagnetic and helimagnetic (FM+HM), helimagnetic (HM) and a paramagnetic phase (PM). The determination of the corresponding transition temperatures is described above. The present experiments suggest that a small amount of In-doping tunes the ground state of the Co$_{3}$Sn$_{2-x}$In$_{x}$S$_{2}$ from FM via a mixed FM + AFM and 
FM + HM states to a HM structure. The critical In-content at which the transition into an incommensurate HM state takes place is $x_{\rm cr,1}$ ${\simeq}$ 0.3. The suppression of the out-of-plane FM state is also supported by the strong reduction of the $c$-axis coercive field, extracted from bulk magnetisation measurements (Fig. 4c). Small $c$-axis coercive field for $x$ ${\geq}$ 0.3 also points to the fact that the moments in the HM state are mostly in-plane. Thus, a helical version of the in-plane AFM state is anticipated at the high In-doping region of the phase diagram. It is important to note that there is a volume wise competition between FM and AFM/HM states, meaning that the FM volume fraction is heavily reduced upon approaching $x_{\rm cr,1}$ (Fig. 4b), while the ordered moment size (Fig. 4b) is only slightly reduced between $x$ = 0 and 0.2 and is abruptly suppressed at $x_{\rm cr,1}$. This demonstrates that the change from FM to HM happens through a first order transition.  Upon further increasing the In-content, the HM ordering temperature $T_{\rm Hel}$ gradually decreases and is suppressed towards the critical doping $x_{\rm cr,2}$ ${\simeq}$ 1, at which the system exhibits a metal to semiconductor transition. Indium doping also causes a smooth reduction of the internal field $B_{int,Hel}$, i.e. the ordered moment size of the HM state, to zero at $x_{\rm cr,2}$ ${\simeq}$ 1 (Fig. 4b), suggesting a continuous QPT across $x_{\rm cr,2}$. Moreover, the magnetically ordered fraction $V_{\rm m}$ is unchanged as the In-content is increased from 0 to 0.9, indicating a fully ordered volume for all samples, exhibiting long-range magnetic order.  $V_{\rm m}$ is reduced drastically for $x_{\rm cr,2}$ ${\simeq}$ 1. For $x$ = 1.1, $V_{\rm m}$ approaches nearly zero. The continuous suppression of the ordered moment size and the sharp reduction of $V_{\rm m}$ from 1 to ${\sim}$ 0 across $x_{\rm cr,2}$ classiffies the transition from long-range magnetic metal to a paramagnetic semiconductor  as a second-order quantum phase transition. This provides the direct evidence of a QCP at $x_{\rm cr,2}$ in Co$_{3}$Sn$_{2-x}$In$_{x}$S$_{2}$. This is to the best of our knowledge the first microscopic experimental observation of a QCP in a topological kagome magnet. The presence of the QCP in Co$_{3}$Sn$_{2-x}$In$_{x}$S$_{2}$ suggests that the physical properties of this material at low temperatures are controlled by the QCP. Indeed, observed \cite{KassemPhd} logarithmic divergence of resistivity near $x$ ${\simeq}$ 1 as the temperature decreases, the observed increased anisotropy of the resistivity and strongly enhanced $T^{3}$ term of the specific heat instead of the linear $T$-term of the specific heat at low temperatures, all indicate an unconventional electronic state of this system near the putative QCP $x$ ${\simeq}$ 1. Our high pressure ${\mu}$SR measurements (see the details in the Supplementary Figure 8) show strong and unusual reduction of the pressure induced suppression rate of the magnetic ordering temperature for the sample $x$ = 0.9 as compared to the suppression rate for the samples $x$ ${\leq}$ 0.8, which might also be related to the proximity of the $x$ = 0.9 sample to the putative quantum critical point at $x$ ${\simeq}$ 1. The presence of small dynamic magnetic regions in the $x$ = 1sample may also be linked to the QCP.

According to our previous experiments and DFT calculations \cite{GuguchiaNat}  the energies of the out-of-plane FM and in-plane AFM interactions are similar, indicating that Co spins have both ferromagnetic interactions along $c$-axis and antiferromagnetic interactions within the kagome plane of Co$_{3}$Sn$_{2}$S$_{2}$. The HM structure in this system may be interpreted by a competition between dominant Heisenberg-type (FM and AFM) interactions and a weaker antisymmetric Dzyaloshinskii-Moriya (DM) interaction. It seems that In-doping affects the competing interactions such that it promotes the HM state. We note that hydrostatic pressure also causes a suppression of both FM and AFM states \cite{XuliangChen}, but a pressure as high as 20 GPa is needed at which both orders are suppressed simultaneously. No HM phase was induced by pressure. In Co$_{3}$Sn$_{2-x}$In$_{x}$S$_{2}$ however, only a small amount of In is sufficient to push the system towards the HM state. Indium substitution introduces holes to the system and at the same time increases the separation of the kagome layers, while hydrostatic pressure shrinks the lattice and no doping is expected. This suggests that lattice expansion and/or hole doping disfavours the out-of-plane FM state and leads to an incommensurate HM state.
The present ${\mu}$SR experiments along with the AHC and $c$-axis magnetic hysteresis measurements suggest the in-plane HM state (moments rotate within the $ab$-plane, see the discussion below) for $x$ ${\geq}$ 0.3. A helical variation of the in-plane AFM structure (see Fig. 4e) is possible. However, additional experiments such as polarized neutron scattering is necessary to determine the precise nature of this novel magnetic phase. The advantage of ${\mu}$SR is its extreme sensitivity to low-moment magnetism as it is the case for Co$_{3}$Sn$_{2}$S$_{2}$.

One of the most striking effects in Co$_{3}$Sn$_{2}$S$_{2}$ is a large intrinsic anomalous Hall conductivity (AHC) ${\sigma}_{xy}$, due to the considerably enhanced Berry curvature arising from its band structure \cite{FelserCSS}. Recently, it was shown that ${\sigma}_{xy}$ features a maximum at $x$ ${\simeq}$ 0.15 above which it sharply decreases and becomes negligibly small for $x$ ${\geq}$ 0.3, well below the magnetic to paramagnetic critical doping $x_{\rm cr,2}$ ${\simeq}$ 1. This indicates that the doping induced disappearance of the ${\sigma}_{xy}$ is correlated with the suppression of the FM state and that around the critical doping $x_{\rm cr,1}$ ${\simeq}$ 0.3, both the ferromagnetism and the AHE are strongly suppressed simultaneously. This can be understood by the first principles calculations, showing that the AHC is dominated by the FM $c$-axis component \cite{GuguchiaNat} (see the supplementary Figure 5). Thus, the data provide direct experimental evidence for the quantum tuning (In-doping) of anomalous Hall conductivity mediated by changes in the competing FM, AFM and HM structures in the kagome lattice. It is remarkable that the phase diagram (Figure 4a) exhibits a doping region where different magnetically ordered phases coexist at zero temperature. The interplay between these regions, each of which may possess distinct topological responses, can possibly give rise to exciting physics at the magnetic domain boundaries. The non-zero but small value of ${\sigma}_{xy}$ and the coercive field for $x$ ${\geq}$ 0.3 (Figure 4c) may indicate either the presence of an experimentally non-resolvable small FM fraction in these high In-doped samples or the presence of some canting within the in-plane HM structure, giving rise to a small $c$-axis net FM moment and a related AHC component. Note that the AHC in the undoped system scales with the FM volume fraction as a function of temperature \cite{GuguchiaNat}. Since the FM fraction is gradually reduced by In-doping, one would expect the gradual suppression of ${\sigma}_{xy}$ as well. Instead, we see a sharp peak of ${\sigma}_{xy}$ at $x$ ${\simeq}$ 0.15. There may be several possible reasons for this behaviour: (1) According to first principle calculations, the anomalous Hall conductivity ${\sigma}_{xy}$ strongly depends on the electron density $n_{\rm e}$ (which does not change much with temperature). In-doping causes the reduction of the FM fraction, but it may modify $n_{\rm e}$ in the FM volumes of the sample in such a way that ${\sigma}_{xy}$ increases. Once the FM fraction becomes negligible, ${\sigma}_{xy}$ also becomes negligibly small. (2) The maximum of ${\sigma}_{xy}$ occurs in the doping range where FM and HM regions coexist. Accordingly, a possible electron/hole transfer between FM and HM phases appears possible, thereby providing a higher effective doping for the FM region and therefore contributing to the increase of the ${\sigma}_{xy}$. (3) Indium doping might modify the band structure leading to an enhancement of ${\sigma}_{xy}$.

\section{Summary}

In summary, the transition metal based kagome material Co$_{3}$Sn$_{2-x}$In$_{x}$S$_{2}$ is an ideal platform to study two intriguing, and in this case intertwined, properties of quantum materials, namely magnetism and topological electronic band structure. Our key finding is the possibility of an effective quantum tuning of the competition between FM, in-plane AFM and mostly in-plane HM orders within the kagome plane of Co$_{3}$Sn$_{2-x}$In$_{x}$S$_{2}$ by In-doping and the presence of multiple quantum phase transitions of different nature in the phase diagram of this system. Our experiments show the presence of at least two zero temperature phase transitions in the phase diagram of this system. One at the low In-doping of $x_{\rm cr,1}$ ${\simeq}$ 0.3, where a QPT from a FM to a HM state through mixed states is observed and a second one at $x_{\rm cr,2}$ ${\simeq}$ 1, with a transition from a helical metallic to a paramagnetic semiconducting state. Remarkably, the QPT from FM to incommensurate HM state is first order, while the transition from HM metal to a PM semiconductor is a continuous/second-order transition. Thus, we have uncovered a quantum critical point (QCP) in Co$_{3}$Sn$_{2-x}$In$_{x}$S$_{2}$ at the doping level of $x_{\rm cr,2}$ ${\simeq}$ 1. The interplay between the competing magnetic states, the charge carrier density within the FM regions and the spin-orbit coupled band structure further seem to induce non-trivial variations of the topological properties of Co$_{3}$Sn$_{2-x}$In$_{x}$S$_{2}$. This is evidenced by a non-monotonous In-doping dependence of the anomalous Hall conductivity with the nearly full suppression of ${\sigma}_{xy}$ above $x_{\rm cr,1}$ ${\simeq}$ 0.3, where the FM state disappears. Therefore the exciting perspective arises of a magnetic system in which the topological response can be controlled, and thus explored, over a wide range of parameters.

\section{METHODS}

\textbf{${\mu}$SR experiment}:

 In a ${\mu}$SR experiment nearly 100${\%}$ spin-polarized muons ${\mu}$$^{+}$
are implanted into the sample one at a time. The positively
charged ${\mu}$$^{+}$ thermalize at interstitial lattice sites, where they
act as magnetic microprobes. In a magnetic material the
muon spin precesses in the local field $B_{\rm \mu}$ at the
muon site with the Larmor frequency ${\omega}$ = 2${\pi}$${\nu}_{\rm \mu}$ = $\gamma_{\rm \mu}$/(2${\pi})$$B_{\rm \mu}$ (muon gyromagnetic ratio $\gamma_{\rm \mu}$/(2${\pi}$) = 135.5 MHz T$^{-1}$).

 The GPS (${\pi}$M3 beamline) \cite{GPS} and the HAL-9500 (${\pi}$E3 beamline) ${\mu}$SR instruments at the Paul Scherrer Institute, Switzerland, were used to study the single crystalline samples of Co$_{3}$Sn$_{2-x}$In$_{x}$S$_{2}$ at ambient pressure. ${\mu}$SR experiments under pressure were performed at the ${\mu}$E1 beamline of the Paul Scherrer Institute (Villigen, Switzerland, where an intense high-momentum ($p_{\mu}$ = 100 MeV/c) beam of muons is implanted in the sample through the pressure cell. The specimen was mounted in a He gas-flow cryostat with the $c$-axis parallel to the muon beam direction and the temperature was varied between 3 and 300 K.\\



\textbf{Analysis of Weak TF-${\mu}$SR data}:

The weak TF asymmetry spectra were analyzed \cite{AndreasSuter} using the function:

\begin{equation}
\begin{aligned}A_S(t)= A_{p}exp(-\lambda t)\cos(\omega t + \phi).
\label{eq1}
\end{aligned}
\end{equation}

where $t$ is the time after muon implantation, $A$($t$) is the time-dependent asymmetry, $A_{p}$
is the amplitude of the oscillating component (related to the paramagnetic volume
fraction), ${\lambda}$ is an exponential damping rate due to paramagnetic spin fluctuations
and/or nuclear dipolar moments, ${\omega}$ = 2${\pi}$${\nu}_{\rm \mu}$ is the Larmor precession frequency set by the
strength of the transverse magnetic field, and ${\phi}$ is a phase offset. The zero for $A$($t$) was allowed to vary for each
temperature to deal with the asymmetry baseline shift occurring for
magnetically ordered samples. From these
refinements, the magnetically ordered volume fraction at each temperature $T$ was
obtained by 1  - $A_{p}(T)$/$A_{p}(T_{max})$, where $A_{p}$($T_{max}$) is the amplitude in the paramagnetic
phase at high temperature.\\

\textbf{Analysis of ZF-${\mu}$SR data}:

The ZF-${\mu}$SR spectra for $x$ ${\textless}$ 0.3 were analyzed using the following commensurate model:
	
	
	\begin{widetext}
		\begin{equation}
		\label{eq:ZFPolarizationFit}
		A_{\textrm{ZF}}(t) = F \left( \sum_{j = 1}^{2}  \left( f_j \cos(2\pi\nu_j t + \phi) e^{-\lambda_j t} \right) +
		f_L e^{-\lambda_{L} t} \right) + (1-F) \left(\frac{1}{3} + \frac{2}{3}\left( 1 - (\sigma t)^2  \right) e^{-\frac{1}{2}(\sigma t)^2} \right)
		\end{equation}
	\end{widetext}
	The model in~\eqref{eq:ZFPolarizationFit} consists of an anisotropic magnetic contribution
	characterized by an oscillating ``transverse'' component and a slowly relaxing ``longitudinal''
	component.  The longitudinal component arises due to muons experiencing local field components which are parallel to the initial muon spin polarization. In polycrystalline samples with randomly oriented fields
	the orientational averaging results in a so-called ``one-third tail'' with $f_L = \frac{1}{3}$. For single crystals, $f_L$
	varies between zero and unity as the orientation between field and polarization changes from being
	perpendicular to parallel. Note that the whole volume of the sample is magnetically ordered nearly up to $T_{\rm C}$. Only very close to the transition, in addition to the magnetically ordered contribution, there is a paramagnetic 
	signal component characterized by the width ${\sigma}$ of the field distribution at the muon site created by the densely distributed nuclear moments. The temperature-dependent magnetic ordering fraction $0
	\le F \le 1$ governs the trade-off between magnetically-ordered and paramagnetic behaviors.\\

    The field distribution for the samples $x$ ${\geq}$ 0.3 presented in Figure 1b is characterized by a minimum ($B_{\rm min}$) and a maximum ($B_{\rm max}$) cutoff field, which is consistent with the incommensurate helimagnetic order, and is well described by the field distribution given by \cite{AmatoPRB,Schenck,Yaouanc,KhasanovPRBrapid,KhasanovJPCM}:

    \begin{equation}  
		P(B) = \frac{2}{\pi}\frac{B}{\sqrt{(B^{2}-B_{min}^{2})(B_{max}^{2}-B^{2})}}
 \end{equation}\\


\textbf{Analysis of ZF-${\mu}$SR data under pressure}:

A substantial fraction of the ${\mu}$SR asymmetry originates
from muons stopping in the MP35N pressure cell surrounding the sample.
Therefore, the ${\mu}$SR data in the whole temperature range were analyzed by
decomposing the signal into a contribution of the sample and a contribution of the pressure cell:
\begin{equation}
A(t)=A_S(0)P_S(t)+A_{PC}(0)P_{PC}(t),
\end{equation}
where $A_{S}$(0) and $A_{PC}$(0) are the initial asymmetries and $P_{S}$(t) and $P_{PC}$(t)
are the time-dependent muon-spin polarizations belonging to the muons stopping in the sample and the pressure cell, respectively. The pressure cell signal was analyzed by a damped Kubo-Toyabe function \cite{Toyabe,GuguchiaNature}.\\


\textbf{Pressure cell}:  Pressures up to 2.1 GPa were generated in a double wall piston-cylinder
type of cell made of CuBe/MP35N material, specially designed to perform ${\mu}$SR experiments under
pressure \cite{GuguchiaPressure,Andreica,GuguchiaNature}. As a pressure transmitting medium Daphne oil was used. The pressure was measured by tracking the superconducting transition of a very small indium plate by AC susceptibility. The filling factor of the pressure cell was maximized such that the fraction of the muons stopping in the sample was approximately 
40${\%}$.\\

\textbf{Data Availability}:

All relevant data are available from the authors. The data can also be found at the following link http://musruser.psi.ch/cgi-bin/SearchDB.cgi using the details: HAL-Year 2019. GPS-Year 2019. GPD-Year 2019.

\section{Acknowledgments}~
The ${\mu}$SR experiments were carried out at the Swiss Muon Source (S${\mu}$S) Paul Scherrer Insitute, Villigen, Switzerland. M.Z.H. acknowledges visiting scientist support from IQIM at the California Institute of Technology. T.N. and S.S.T. acknowledge funding from the European Research Council (ERC) under the European Unions Horizon 2020 research and innovation programm (ERC-STG-Neupert-757867-PARATOP), and also from NCRR Marvel. \\


\textbf{Competing interests:} The authors declare that they have no competing interests.\\

\newpage

\begin{center}
\textbf{Supplementary information}
\end{center}

\subsection{Pressure cell}  Pressures up to 2.1 GPa were generated in a double wall piston-cylinder
type of cell made of CuBe/MP35N material, especially designed to perform ${\mu}$SR experiments under
pressure \cite{GuguchiaPressure,Andreica,GuguchiaNature}. As a pressure transmitting medium Daphne oil was used. The pressure was measured by tracking the superconducting transition of a very small indium plate by AC susceptibility. The filling factor of the pressure cell was maximized such that the fraction of the muons stopping in the sample was approximately 
40${\%}$.\\

\subsection{Calculation of anomalous hall conductivity for various magnetic configurations for Co$_{3}$Sn$_{2}$S$_{2}$}

For the system Co$_{3}$Sn$_{2}$S$_{2}$ We considered maximally symmetric Shubnikov two magnetic subgroups, as discussed in our previous work \cite{GuguchiaNat}. The first subgroup R-3m$'$ has two spin components $\bf{M}_{\rm Co}$ = ($m_{\rm x}$,2$m_{\rm x}$,$m_{\rm z}$) with AFM in the plane and FM along the $c$-axis (Supplementary Figure 5). The second subgroup R-3m is 120$^{\circ}$ antiferromagnetic (AFM) order with the spins in ($ab$) plane (see Fig. 3e), which we will refer to as in-plane AFM order (Supplementary Figure 5c). According to density functional theory (DFT) calculations, among the two magnetic arrangements as shown in Supp. Fig. 5b and c, the lowest energy configuration is FM with the magnetic moments along the $c$-axis, with magnitude ${\sim}$ 0.35 ${\mu_B}$/atom. The in-plane AFM configuration with the same magnetic moment has a higher energy of ${\sim}$ 20 meV/Co atom, relative to the $c$-axis FM order. Supplementary Figure 5a shows the dependence of the in-plane anomalous hall conductivity ${\sigma}_{xy}$ on the chemical potential, calculated for above mentioned magnetic structures. As one can see for the fully $c$-axis oriented R-3m$'$ FM ordering, we obtain an extremely high ${\sigma}_{xy}$ = 10$^{3}$ ${\Omega}$$^{-1}$cm$^{-1}$. When the spins are fully canted into the plane within R-3m$'$ subgroup, ${\sigma}_{xy}$ acquires much smaller value. The in-plane AFM R-3m structure, shown in Supplementary Fig. 5c, has a 2-fold rotational symmetry with an axis in the $ab$ plane. Such a rotation transforms ${\sigma}_{xy}$ to -${\sigma}_{xy}$, and hence forces it to be completely zero. Thus, one concludes from first principles calculations that the AHC is dominated by the FM $c$-axis component.

\subsection{Dominant paramagnetic response in the Co$_{3}$SnInS$_{2}$ sample.}

Supplementary Fig. 6 displays representative zero-field (ZF) ${\mu}$SR time spectra for Co$_{3}$Sn$_{2-x}$In$_{x}$S$_{2}$
with $x$ = 1 taken at 5 K and 7.5 K. At $T$ = 7.5 K, the entire sample is in the paramagnetic state as evidenced by the weak ${\mu}$SR depolarization and its Gaussian functional form arising from the interaction between the muon spin and randomly oriented nuclear magnetic moments \cite{Toyabe}. At lower temperatures (see spectrum at 5 K in sup. Fig. 2), we observe relaxation occurring only in a small fraction of the signal with no well-defined oscillations, indicating the existence of disordered static magnetism or dynamic fluctuations in a small volume fraction of the sample. These measurements indicate that the $x$ = 1 sample has dominant paramagnetic response down to the lowest temperature measured.  

A fast depolarization of the implanted muons in a small volume fraction in $x$ = 1 sample was found to arise from the fluctuating magnetic moments. That the fluctuations are the origin of the observed muon spin depolarization is evidenced by measurements under 100 mT, which decouples the paramagnetic response in the large $t$-range, but shows no field effect on the initial strongly damped signal (see Figure 7). If the fast depolarization were caused by a wide distribution of static fields, then a small field would be enough to fully recover the muon spin polarization. The presence of small slowly fluctuating magnetic regions in the $x$ = 1 sample may be linked to the quantum critical point. It may also point to the tendency of this kagome system towards the spin-liquid state.

\subsection{Pressure effects on magnetic volume fraction and the magnetic transition temperature in Co$_{3}$Sn$_{2-x}$In$_{x}$S$_{2}$}

 For further insight into the magnetic order in Co$_{3}$Sn$_{2-x}$In$_{x}$S$_{2}$, weak-TF ${\mu}$SR experiments were carried out as a function of hydrostatic pressure for the samples $x$ = 0.8 and 0.9. These values of $x$ place the system near the putative QPT. The temperature dependence of the magnetically ordered volume fraction for these samples, recorded at various applied pressures, are shown in Supplementary Figure 8a. Supp. Figures 8b and 8c depict the magnetic transition temperatures of the samples, normalised to the value at the ambient pressure, and the absolute value of the pressure induced shift of $T_{\rm C2}$, respectively, as a function of the hydrostatic pressure. Two important conclusions can be drawn from these data: (1) While pressure causes the suppression of the magnetic ordering temperature and thus moves the system very close to a QPT, the magnetic volume fraction stays nearly 100 ${\%}$ for the entire pressure range. This suggests that pressure has the tendency to drive the system towards a second-order quantum phase transition, similar to the In-doping. (2) The critical pressure (see Supp. Fig. 8b)  is significantly reduced between $x$ = 0 with $T_{\rm C2}$ ${\simeq}$ 172 K and 0.8 with $T_{\rm C2}$ ${\simeq}$ 43 K samples, as expected. This is understood by the same absolute value (see Supp. Fig. 8c) of the shift ${\Delta}$$T_{\rm C2,onset}$ = $T_{\rm C2,onset}$($p$) - $T_{\rm C2,onset}$(0) for $x$ = 0 and 0.8 samples. However, the shift ${\Delta}$$T_{\rm C2,onset}$ is significantly reduced for $x$ = 0.9 sample (see Supp. Fig. 8c and its inset) with ambient pressure $T_{\rm C2}$ ${\simeq}$ 15 K, giving rise to the same critical pressure for $x$ = 0.9 as for the $x$ = 0.8 sample (see Supp. Fig. 8b). This is unusual since one expects the faster suppression of magnetic order by pressure for $x$ = 0.9 than for $x$ = 0.8 since the former is located closer to QPT and have much lower $T_{\rm C2}$. The departure from the standard behaviour is very interesting and might be related to the proximity of the $x$ = 0.9 sample to the putative quantum critical point at $x$ = 1.










\begin{figure*}[t!]
\includegraphics[width=1.0\linewidth]{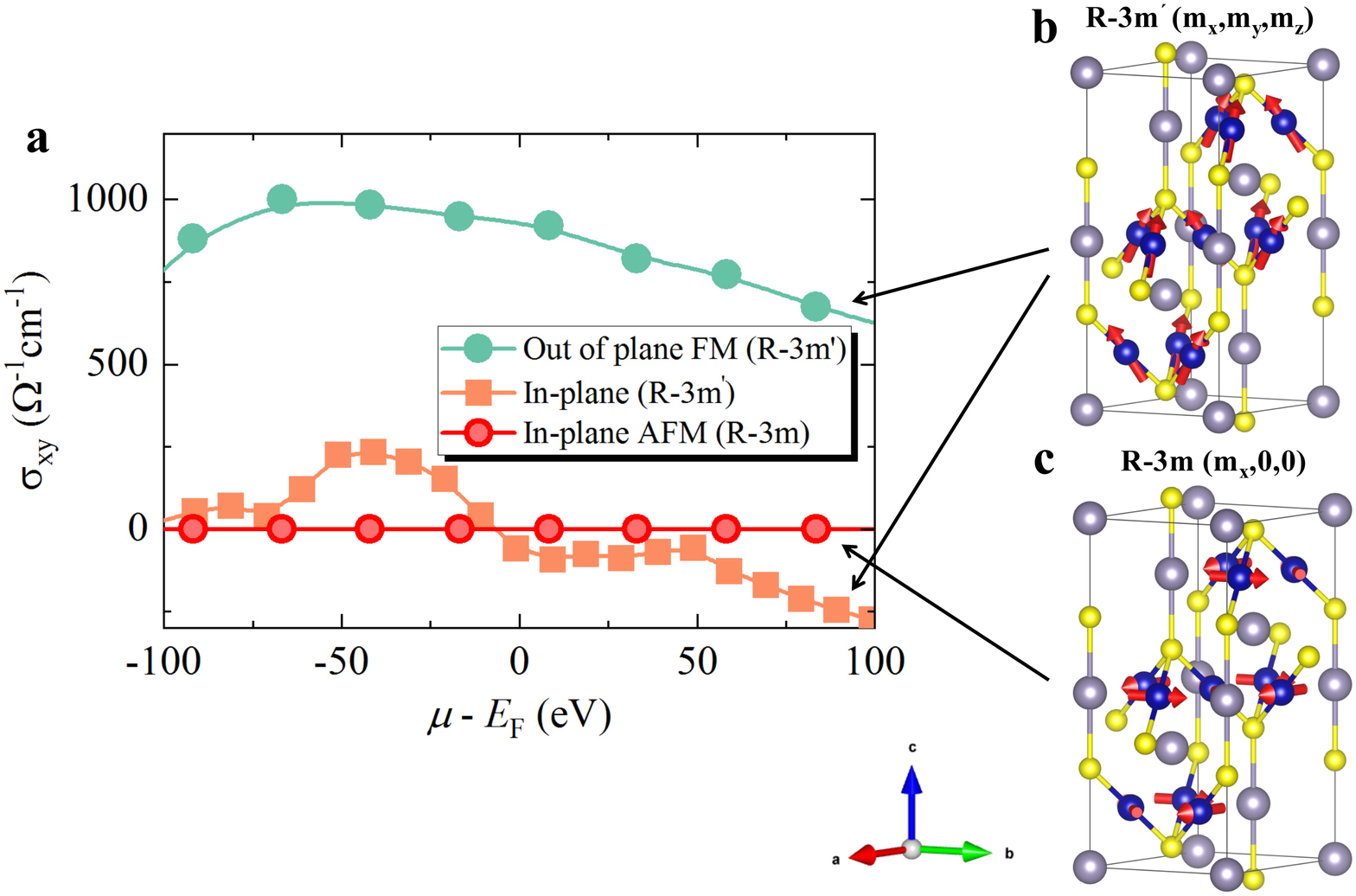}
\vspace{-1.0cm}
\caption{ (Color online) \textbf{Calculated anomalous hall conductivity.}
(a) The dependence of ${\sigma}_{xy}$ on the chemical potential, calculated for the R-3m$'$ and R-3m magnetic structures. 
(b) R-3m$'$ magnetic structure that has two spin components $\bf{M}_{\rm Co}$ = ($m_{\rm x}$,2$m_{\rm x}$,$m_{\rm z}$) with AFM in the plane and FM along the $c$-axis. (c) In-plane R-3m structure, which is 120$^{\circ}$ antiferromagnetic (AFM) order with the spins in ($ab$) plane. Gray solid lines indicate the boundaries of a single unit cell of the crystal structure.}
\label{fig1}
\end{figure*}
%
\begin{figure*}[t!]
\includegraphics[width=0.8\linewidth]{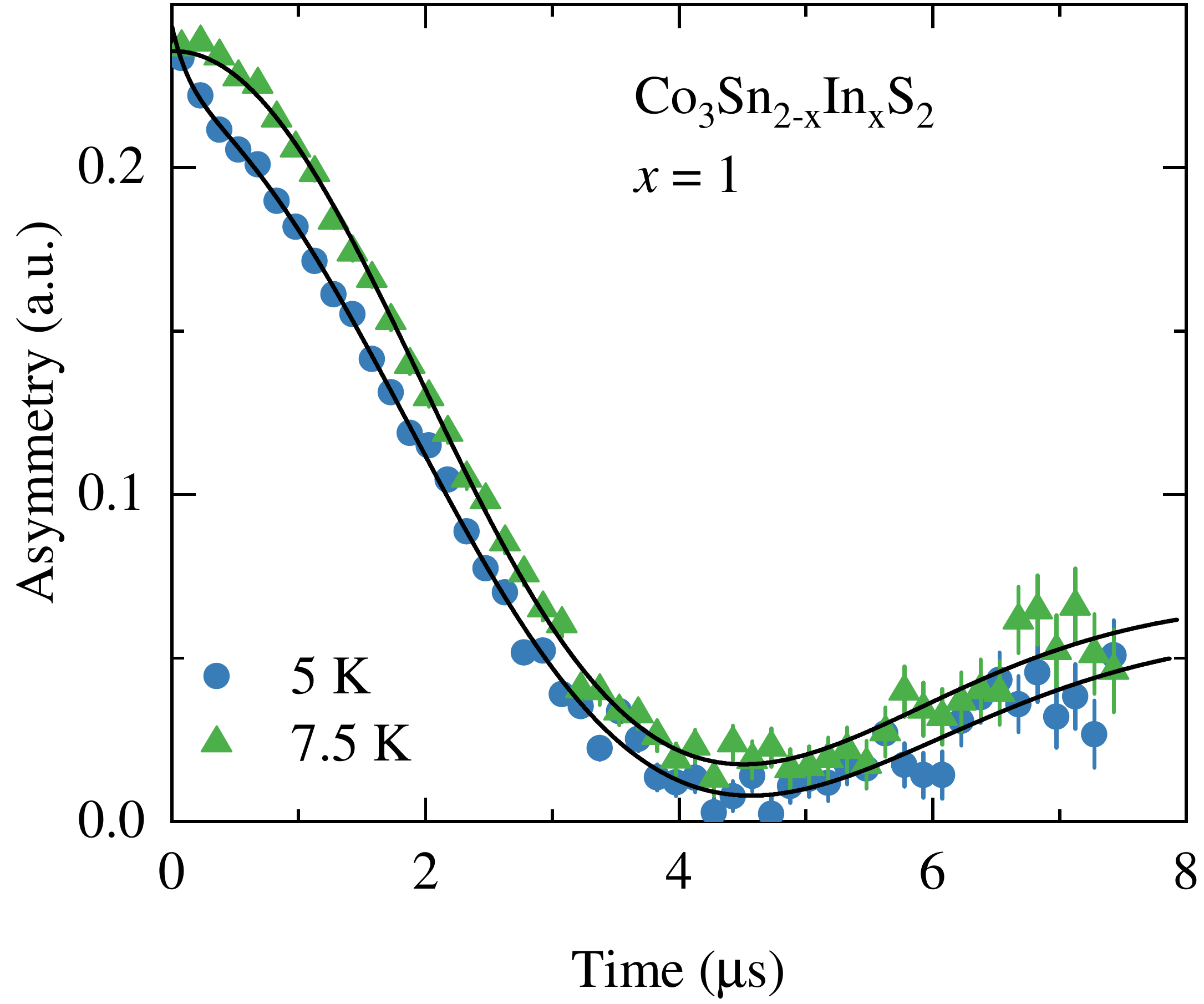}
\vspace{0cm}
\caption{ (Color online) \textbf{Dominant paramagnetic response in CoSnInS$_{2}$.}
Zero-field (ZF) ${\mu}$SR time spectra for the sample $x$ = 1, recorded at 5 K and 7.5 K.}
\label{fig1}
\end{figure*}

\begin{figure*}[t!]
\includegraphics[width=0.8\linewidth]{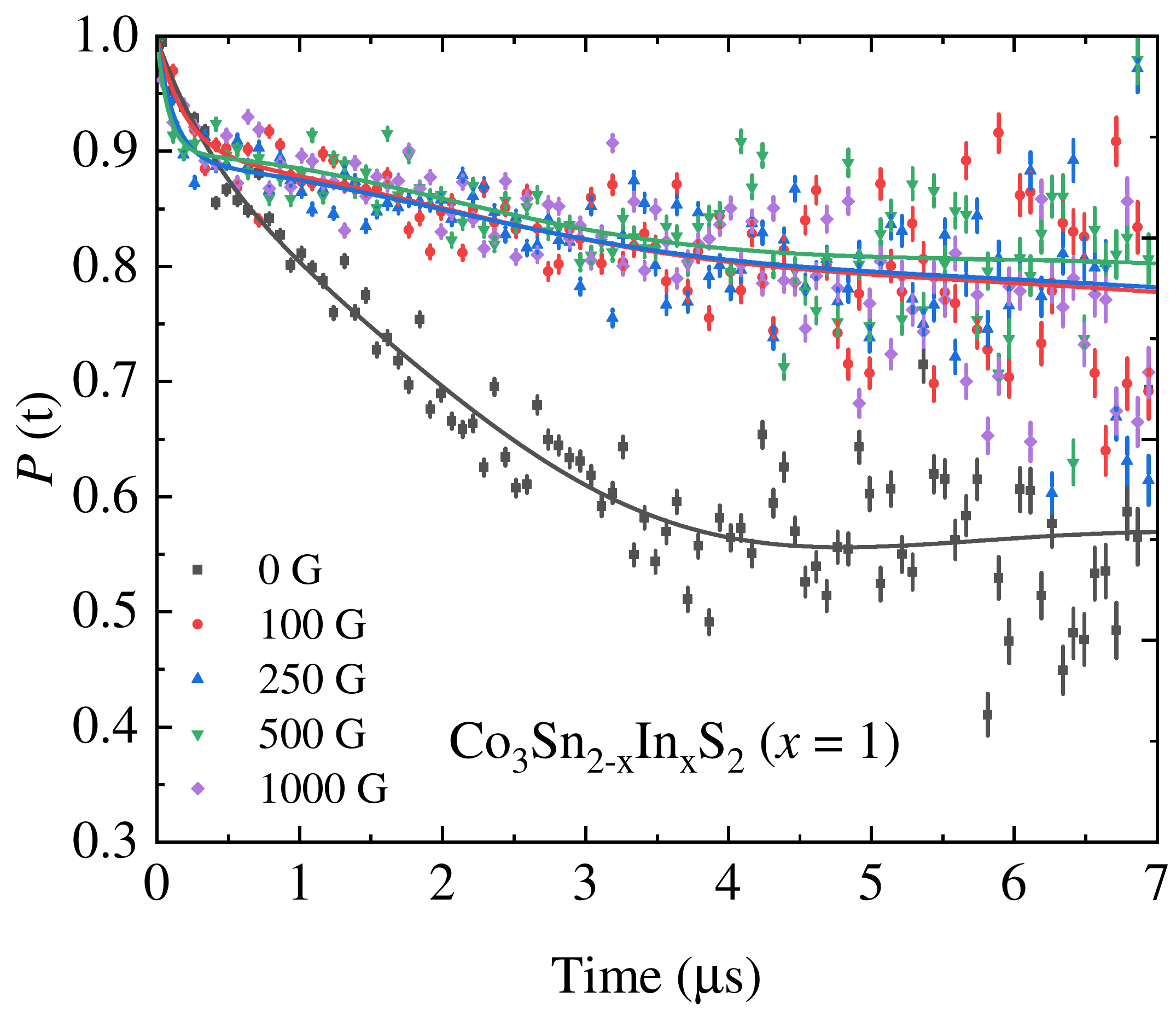}
\vspace{0cm}
\caption{ (Color online) \textbf{Dynamic response from the small volume of the sample in CoSnInS$_{2}$.}
Zero-field (ZF) and longitudinal-field ${\mu}$SR time spectra for the sample $x$ = 1, recorded at 2 K.}
\label{fig1}
\end{figure*}

\begin{figure*}[t!]
\includegraphics[width=0.45\linewidth]{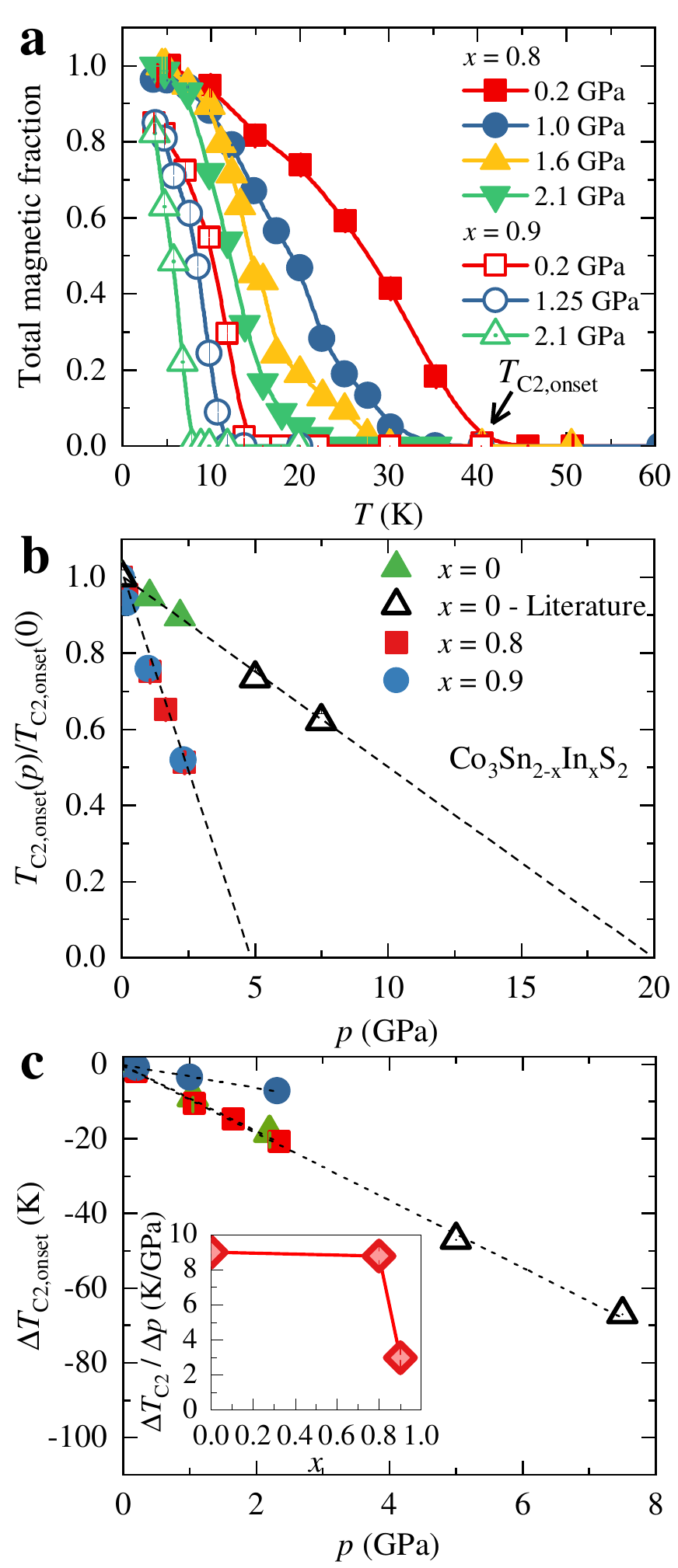}
\vspace{0cm}
\caption{ (Color online) \textbf{Pressure effects on magnetism in CoSn$_{2-x}$In$_{x}$S$_{2}$.}
(a) The temperature dependence of the total magnetic volume fraction for $x$ = 0.8 and 0.9 samples, shown for various hydrostatic pressures. (b) The pressure dependence of $T_{\rm C2}$, normalised to its ambient pressure value,  for $x$ = 0 \cite{GuguchiaNat}, 0.8 and 0.9 samples. The dashed lines are linear fits to the data. The linear suppression of the critical temperature is taken following the previous work \cite{XuliangChen}. (c) The pressure dependence of the pressure induced absolute shift in critical temperature ${\Delta}$$T_{\rm C2,onset}$ = $T_{\rm C2,onset}$($p$) - $T_{\rm C2,onset}$(0). Inset shows the rate of suppression of the magnetic order ${\Delta}$$T_{\rm C2}$/${\Delta}p$  as a function of In-doping.}
\label{fig1}
\end{figure*}

\end{document}